\documentclass{article}
\usepackage{graphicx}

\begin{document}

\title{Conformally flat, quasi-circular numerical simulations of the gravitational wave chirp from binary neutron star merger GW170817}
\author{Grant J. Mathews$^1$, In-Saeng Suh$^{1,2}$, N. Q. Lan$^1$, Atul Kedia$^1$\\
\small $^1$Center for Astrophysics, Department of Physics, University of Notre Dame,\\
\small Notre Dame, IN 46556, USA\\
\small $^2$Center for Research Computing, University of Notre Dame,\\
\small Notre Dame, IN 46556, USA\\
\small E-mail: gmathews@nd.edu}
\date{}
\maketitle

\begin{abstract}
The first detection of gravitational waves from the binary neutron star merger GW170817 by the LIGO-Virgo Collaboration has provided fundamental new insights into the astrophysical site for r-process nucleosynthesis and on the nature of dense neutron-star matter. The detected gravitational wave signal depends upon the tidal distortion of the neutron stars as they approach merger. We report on relativistic numerical simulations of the approach to binary merger in the conformally flat, quasi-circular orbit approximation. We show that this event serves as a calibration to the quasi-circular approximation and a confirmation of the validity of the conformally flat approximation to the three-metric. We then examine how the detected chirp depends upon the adopted equation of state. This establishes a new efficient means to constrain the nuclear equation of state in binary neutron star mergers.

\end{abstract}


On August 17, 2017, Nature revealed itself in a most spectacular way, both from the gravitational waves detected by the LIGO and VIRGO collaborations \cite{LIGO}, and by the multitude of follow up observations \cite{GRB} of the GRB170817A kilonova, along with IR and optical ground-based observations. In this work we consider two aspects of what may have been learned from this event. On the one hand, the GW chirp implies a possible dilemma regarding the nuclear equation of state (EoS) deduced from the post-Newtonian tidal-polarizability vs. observed neutron star properties. Here we discuss our analysis of the chirp in numerical general relativity\cite{Suh17} as a means to better constrain the EoS and clarify the current dilemma. The second aspect considered here is that the kilonova EM spectrum indicates evidence of r-process nucleosynthesis that has implications for the fission barriers and termination of fission recycling during the r-process.

In the LIGO analysis\cite{LIGO} and in follow up analysis \cite{Horowitz,LIGO2} the tidal polarizability was deduced from post-Newtonian expansion. The tidal polarizability (or deformability) $\Lambda$ is an intrinsic neutron-star property highly sensitive to the compactness parameter [$\Lambda \sim (R/M)^5$] that describes the tendency of a neutron star to develop a mass quadrupole as a response to the tidal field induced by its companion \cite{Damour,Hinderer}. Based upon the most recent analysis by LIGO \cite{LIGO2} it has been deduced that the reduced tidal polarizability is $\tilde \Lambda = 190^{+390}_{-120} $ implying that the radius of the stars of $1.4 $ M$_\odot$ is in the range $10.5$ km $\le R \le 13.3 $ km.

However, in this work we consider the possibility that as the stars approach merger during the chirp they are neither well described by post-Newtonian physics, nor as a simple quadrupole deformation. Indeed, one desires to trace the evolution of the binary from when they first enter the LIGO window until some 30 sec later when the stars merge. For an orbit period of $\sim 100$ ms, this would require evolving the system for $\sim 3000$ orbits. This is indeed a daunting task from the standpoint of numerical relativity for which it is difficult to follow more than a few orbits due to limitations of computational resource.

Here, we report on an analysis of a general relativistic hydrodynamic simulation \cite{Suh17,Lan15} that is capable of stably numerically integrating thousands of orbits. The solution of the field equations and hydrodynamic equations of motion is summarized in \cite{wmbook,mw97} based upon the conformally-flat condition on the spatial three-metric. In this approach, one begins with the slicing of spacetime into the usual one-parameter family of hypersurfaces separated by differential displacements in a time-like coordinate as defined in the (3+1) ADM formalism \cite{adm,york79}.

In Cartesian $x, y, z$ isotropic coordinates, proper distance is expressed as
\begin{equation}
\label{conftrans}
ds^2 = -(\alpha^2 - \beta_i\beta^i) dt^2 + 2 \beta_i dx^i dt +
\phi^4\delta_{ij}dx^i dx^j ~~,
\end{equation}
where the lapse function $\alpha$ describes the differential lapse of proper time between two hypersurfaces. The quantity $\beta_i$ is the shift vector denoting the shift in space-like coordinates between hypersurfaces. The curvature of the metric of the 3-geometry is then described by the conformally flat approximation (CFA), i.e. the spatial metric is described described by a position dependent conformal factor $\phi^4$ times a flat-space Kronecker delta ($\gamma_{ij} = \phi^4 \delta_{ij}$). This conformally flat condition on the metric provides a numerically valid initial solution to the Einstein equations. The vanishing of the Weyl tensor for a stationary system in three spatial dimensions guarantees that a conformally flat solution to the Einstein equations exists.

To solve for the fluid motion of the system in curved spacetime it is convenient to use an Eulerian fluid description \cite{wmbook}. By introducing the usual set of Lorentz contracted state variables it is possible to write the relativistic hydrodynamic equations in a form which is reminiscent of their Newtonian counterparts \cite{wmbook}. The hydrodynamic state variables are: the coordinate baryon mass density, $D$, the internal energy density $E$ and the covariant spatial momentum density, $S_i$;
\begin{equation}
D = W \rho~~;~~E = W \rho \epsilon~~;~~S_i = (D + E + PW) U_i~~,
\end{equation}
where $W$ is a Lorentz-like factor. Key to maintaining numerical stability is to evolve the the spatial three velocity in the rotating frame,
\begin{equation}
V^i = \alpha \frac{U_i}{\phi^4 W} - \beta^i~~;
\label{three-vel}
\end{equation}

In terms of these state variables, the hydrodynamic equations in the CFA are as follows: The equation for the conservation of baryon number takes the form,
\begin{equation}
\frac{\partial D}{\partial t} = -6D \frac{\partial \log\phi}{\partial t}
- \frac{1}{\phi^6} \frac{\partial}{\partial x^j}(\phi^6DV^j)~~.\
\end{equation}
The equation for internal energy evolution becomes,
\begin{eqnarray}
\frac{\partial E}{\partial t} =&& -6(E + PW) \frac{\partial \log\phi}{\partial t}
- \frac{1}{\phi^6} \frac{\partial}{\partial x^j}(\phi^6EV^j)\nonumber \\
&& - P\biggl[\frac{\partial W}{\partial t} +
\frac{1}{\phi^6} \frac{\partial}{\partial x^j}(\phi^6 W V^j)\biggr]~~.
\end{eqnarray}
Momentum conservation takes the form,
\begin{eqnarray}
\frac{\partial S_i}{\partial t}& = & -6 S_i \frac{\partial \log\phi}{\partial t}
- \frac{1}{\phi^6} \frac{\partial}{\partial x^j}(\phi^6S_iV^j)
-\alpha \frac{\partial P}{\partial x^i} \nonumber \\
& + & 2\alpha (D+ E + PW)( W - \frac{1}{W}) \frac{\partial \log\phi}{\partial x^i}
+ S_j \frac{\partial \beta^j}{\partial x^i} \nonumber \\
& - & W (D + E + PW) \frac{\partial \alpha}{\partial x^i} - \alpha W
(D+ E + PW) {\partial \chi \over \partial x^i}~~, 
\label{hydromom}
\end{eqnarray}
where the last term in Eq. (\ref{hydromom}) is the contribution from the radiation reaction potential $\chi$ as defined in Refs.~\cite{Suh17,Lan15,wmbook}. Including this term allows for a calculation of the orbital evolution via gravitational wave emission in the CFA.

As a first test and calibration of the CFA we have adopted the series quasi-stable orbits from \cite{Suh17,Lan15} for which gravitational radiation loss is set to zero. The relative time of each orbit could be determined from the timescale for orbital energy loss by gravitational wave emission as described in \cite{Suh17,Lan15},
\begin{equation}
t = t_0 + E/\dot E~~,
\end{equation}
Where $E$ is the orbital energy and $\dot E$ denotes the coordinate time derivative.

The entire chirp could then by constructed by fitting the numerical results for frequency vs. time with a modified power-law form of the chirp given by,
\begin{equation}
\frac{96}{5} \pi^{8/3} {\cal M}^{5/3} t + \frac{3}{8} f^{5/3(a t^2 + b t + c)} + C = 0 ~~.
\end{equation}
Here $a$, $b$ and $c$ are deduced from fits to the numerical simulation and correspond to corrections due to the inclusion of higher moments in density, gravity, mass-energy and momentum in the chirp. This correction is motivated by the observation that the deviation of a logarithmic $f$ vs. $t$ plot from a straight line in numerical simulations is well represented by a simple quadratic curve.

A caveat of this approach, however, is that the time zero ($t_0$) for each series of orbits is not well defined. For our purposes we fixed the time zero as the point at which the slope of the chirp ($df/dt$) matches the slope of the LIGO observation. Since the slope is small, however, this introduces about a 10\% uncertainty in the time-zero calibration. Nevertheless, as we show below, the differences in among the equations of state can be so great, that this is an acceptable uncertainty.

\begin{figure}[t]
\begin{center}
\includegraphics[width=5 in]{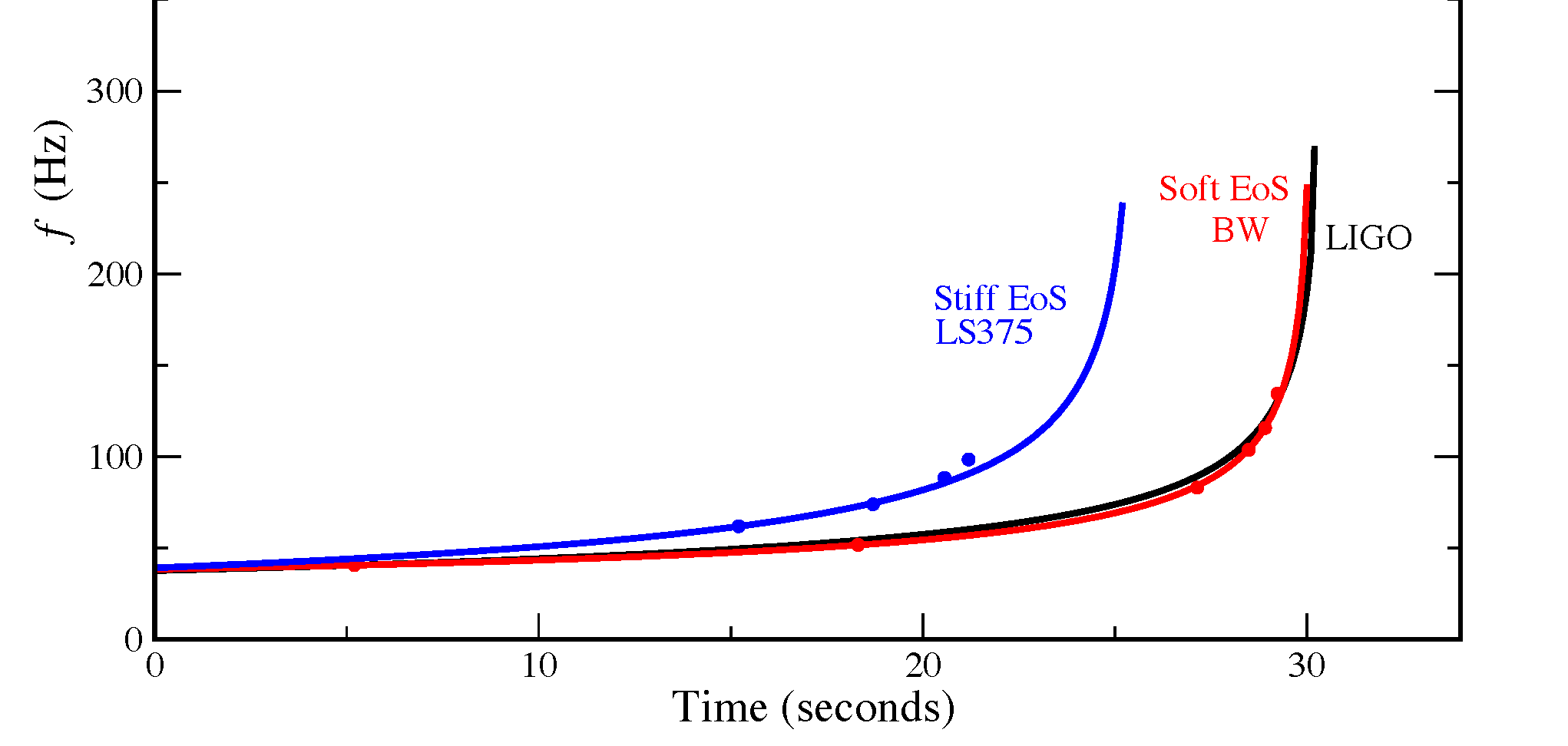}
\end{center}
\caption{Reconstructed LIGO gravitational wave chirp compared with the chirp calculated here based upon two equations of state for neutron stars as labeled.}
\label{fig:1}
\end{figure}

An example of the reconstructed chirp for two different equations of state is shown in figure \ref{fig:1} based upon orbits of two nearly equal mass 1.4 M$_\odot$ neutron stars described in \cite{Suh17,Lan15}. The curve labeled BW is for the soft EoS of Bowers and Wilson \cite{wmbook} which leads to a compact neutron star. The curve labelled LS375 is based upon the extremely stiff, high incompressibility EoS from Lattimer and Swesty \cite{LS91} for which a large neutron-star radius is implied. These extreme examples show a strong sensitivity to the nuclear EoS and is consistent with the sensitivity deduced from the PN analysis of tidal polarizability \cite{LIGO,Horowitz}.

This figure also establishes that the CFA in the quasi-circular orbit approximation can be used to reconstruct the chirp in a much more economical way than to attempt an exact fully relativistic simulation, while also including higher order corrections than the standard PN analysis.

\section*{Acknowledgment}
This work was supported by the U.S. Department of Energy under grant DE-FG02-95-ER40934.


\begin{thebibliography}{9}
\bibitem{LIGO}B. Abbott et al. (LIGO Scientific and Virgo Collaboration), Phys. Rev. Lett. 119, 161101 (2017).
\bibitem{GRB}B. P. Abbott et al. (Virgo, Fermi-GBM, INTEGRAL, and LIGO Scientific Collaboration), Astrophys. J. 848, L12 (2017).
\bibitem{Suh17} I.-S. Suh, G. J. Mathews, J. R. Haywood, and N. Q. Lan, Adv. in Astr., 2017, 612703, (2017) arXiv:1601.01460.
\bibitem{Horowitz}F. J. Fattoyev, J. Piekarewicz, and C. J. Horowitz, Phys. Rev Lett. 120, 172702 (2018).
\bibitem{LIGO2} B. Abbott et al. (LIGO Scientific and Virgo Collaboration), Phys. Rev. Lett. 121, 161101 (2018).
\bibitem{Damour}T. Damour, M. Soffel, and C. Xu, Phys. Rev. D 45, 1017 (1992).
\bibitem{Hinderer} E. E. Flanagan and T. Hinderer, Phys. Rev. D 77, 021502 (2008).
\bibitem{Lan15} N. Q. Lan., I.-S. Suh, G. J. Mathews, and J. R. Haywood, Comm. in Phys., 25, 299, (2015).
\bibitem{wmbook} Wilson J R and Mathews G J {\it Relativistic Numerical Hydrodynamics}, 
(Cambridge University Press, Cambridge, United Kingdom) (2003).
\bibitem{mw97} G. J. Mathews and J. R. Wilson J R, {\it Astrophys. J.} {\bf 482} 929 (1997).
\bibitem{adm} R. Arnowitt, S. Deser S, and C. W. Misner {\it Gen. Rel. Grav.} {\bf 40} 1997 (2008).
\bibitem{york79} J. W. York, Jr in {\it Sources of Gravitational Radiation}, L. L. Smarr, Ed. (Cambridge University Press, Cambridge, UK) p. 83 (1979).
\bibitem{LS91} J. Lattimer and F. D. Swesty Nucl. Phys. A, {\bf 535} 331 (1991).


\end{thebibliography}
\end{document}